\documentclass[doublecol]{epl2}
\usepackage{color}
\usepackage{graphicx}
\usepackage{amssymb}
\usepackage{amsmath}
\usepackage{amsfonts}
\usepackage[utf8]{inputenc}
\usepackage[normalem]{ulem}
\usepackage{times}
\usepackage{float}
\usepackage{epstopdf}
\usepackage{subcaption}

\newcommand{\Tg}{T_{\text g}}
\newcommand{\Tc}{T_{\text c}}

\newcommand{\tauLJ}{\tau_\mathrm{LJ}}

\newcommand{\phig}{\phi_{\text g}}
\newcommand{\rhoA}{\rho_{\text A}}
\newcommand{\rhoB}{\rho_{\text B}}

\newcommand{\vs}{v_{\text {s}}}

\newcommand{\diff}{\text{d}}
\newcommand{\myeq}{\!=\!}
\renewcommand{\vec}{\mathbf}

\newcommand{\kB}{k_{\text {B}}}
\newcommand{\Cs}{C^{\text{s}}}

\newcommand{\rcab}{r_{\mathrm{c},\alpha\beta}}

\newcommand{\Cexz}{\ensuremath{C_{\varepsilon_{xz}}}}
\newcommand{\Cexy}{\ensuremath{C_{\varepsilon_{xy}}}}
\newcommand{\Ceyz}{\ensuremath{C_{\varepsilon_{yz}}}}

\newcommand{\exzna}{\ensuremath{\varepsilon_{xz}}}

\newcommand{\delete}[1]{}

\definecolor{pastell}{rgb}{0.066,0.289,0.519}
\definecolor{darkred}{rgb}{0.585,0,0}
\definecolor{darkgreen}{rgb}{0,0.585,0}
\definecolor{darkblue}{rgb}{0,0,0.585}
\definecolor{lightmagenta}{rgb}{1,0,1}
\definecolor{velvet}{rgb}{0.468,0.13,0.4}
\definecolor{redblue}{rgb}{0.5,0,0.5}
\definecolor{redgreen}{rgb}{0.5,0.5,0}
\definecolor{greenblue}{rgb}{0,0.5,0.5}
\definecolor{maroon}{cmyk}{0,0.87,0.68,0.32}

\newcommand{\hide}[1]{\textcolor{blue}{}}

\usepackage{hyperref}

\title{Long-range strain correlations in 3D quiescent glass forming liquids}
\author{Muhammad Hassani\inst{1}, Elias M.~Zirdehi\inst{1}, Kris Kok\inst{2}, Peter Schall\inst{2}, Matthias Fuchs\inst{3}, Fathollah Varnik\inst{1}
\thanks{Corresponding author: \email {fathollah.varnik@rub.de}}}
\shortauthor{Hassani \etal}
\shorttitle{Strain-fluctuations in 3D glass formers}
\institute{
	\inst{1} ICAMS, Ruhr-Universit\"at Bochum, Universit\"atsstra{\ss}e 150,
	44780 Bochum, Germany\\ 	
	\inst{2} Institute of Physics, University of Amsterdam, Science Park 904, 1098XH Amsterdam, The Netherlands\\
	\inst{3} Fachbereich Physik, Universität Konstanz, 78457 Konstanz, Germany
}

\abstract{
We present a quantitative study of strain correlations in quiescent supercooled liquids and glasses. Recent two-dimensional computer simulations and experiments indicate that even supercooled liquids exhibit long-lived, long-range strain correlations. Here we investigate this issue in three dimensions via experiments on hard sphere colloids and molecular dynamics simulations of a glass forming binary Lennard Jones mixture. Both in the glassy state and in the supercooled regime, strain correlations are found to decay with a $1/r^3$ power-law behavior, reminiscent of elastic fields around an inclusion. Moreover, theoretical predictions on the time dependence of the correlation amplitude are in line with the results obtained from experiments and simulations.  It is argued that the size of the domain, which exhibits a 
cooperative strain pattern in a supercooled liquid, is determined by the product of the speed of sound with the structural relaxation time. While this length is of the order of  nanometers in the normal liquid state, it grows to macroscale when approaching the glass transition.
}

\begin{document}

\maketitle
	
\section{Introduction}
The existence of a growing length scale upon approaching the glass transition has been the subject of intense studies (see, e.g.,~\cite{Bennemann1999a,Karmakar2016} and references therein). While earlier works mostly addressed quiescent systems, an alternative perspective to study glassy behavior has emerged in the last decade with a focus on non-equilibrium response~\cite{Berthier2000,Fuchs2002}. Under steady shear, glasses show evidence of long-range strain correlations, resulting from the elastic coupling of local shear transformation zones~\cite{Falk1998}. This leads to a strongly correlated strain pattern which resembles the Eshelby solution around a pre-sheared spherical inclusion in a homogeneous isotropic elastic medium \cite{picard2004,Chikkadi2011,Nicolas2014,Talamali2012,Hassani2018}. Relevance of these correlations for shear banding is also discussed in the literature \cite{Chikkadi2011,Dasgupta2013,Chikkadi2012b,Mandal2013,Chikkadi2014,Hassani2016}.
Interestingly, long-range Eshelby-type correlations of strain fluctuations have  been  reported in molecular dynamics simulations of a quiescent supercooled liquid in two dimensions \cite{Chattoraj2013}. Evidence for a four-fold symmetry of strain correlations has also been found from experiments on a quiescent colloidal glass~\cite{Jensen2014}. These observations have been recently rationalized in terms of the non-equilibrium Mode Coupling Theory (MCT), predicting additionally the time dependence of the correlation amplitude both in the glassy phase and in the supercooled regime~\cite{Illing2016}. These predictions have been tested via two-dimensional Brownian dynamics simulations of hard discs and experiments on colloidal mono-layers~\cite{Illing2016}.

These long-range spatial correlations should play an important role in understanding the divergence of the structural relaxation time from supercooled liquid to glass. Yet, despite the fact that dimensionality plays a major role in the elastic response of solids---Eshelby-fields decay as $1/r^d$ with $d$ being the spatial dimension---a detailed quantitative study of this problem in three dimensions is still lacking (see, however,~\cite{Jensen2014} for a first study in the glassy state). The present work focuses on this point via a combination of confocal microscopy experiments on hard-sphere colloids and molecular dynamics (MD) simulations of a binary Lennard-Jones (LJ) model, well-known for its glass forming property. In line with theoretical predictions, the autocorrelation function of shear strain fluctuations is shown to decay as $1/r^3$  both in the glassy state and in the supercooled regime. When plotted versus the time interval used to evaluate strain fluctuations, the amplitude of these correlations reaches a plateau in the glassy state but grows linearly in the supercooled regime. A discussion of this behavior in terms of a finite-sized elastic domain around a shear deformation event is presented.

\section{Experiments}

To measure strain correlations in an experimental system, we use hard-sphere colloidal suspensions that are good model systems for glasses. In these systems, structural relaxation slows down beyond bounds at particle volume fractions larger than $\phig \sim 0.58$, the colloidal glass transition~\cite{vanMegen1994}. We use sterically stabilized fluorescent polymethylmethacrylate (PMMA) particles with a diameter of $a = 2R = 1.47 \mu$m and a polydispersity of $6\%$ to prevent crystallization, suspended in a density and refractive-index matching mixture of cyclohexyl bromide and cis-decalin. Care was taken to meticulously balance the solvent composition to achieve best density match after particles had swollen to their final size. To screen any possible residual charges, we also added a small amount of the organic salt TBAB (tetrabuthyl ammonium bromide). Suspensions at volume fractions of $\phi =$ 0.32, 0.56 and 0.6 were then prepared by diluting samples quenched to a sediment, assuming a volume fraction of the sediment of $64 \%$; the quench was achieved by changing the temperature to $29^\circ C$, above room temperature, where due to differences in thermal expansion the particles become slightly heavier than the solvent.
We image individual particles in three dimensions using confocal microscopy, and determine their positions with an accuracy of $40$nm in the vertical, and $20$nm in the horizontal direction using an iterative tracking algorithm. Three-dimensional image stacks were acquired every minute over a time interval of 20 minutes. Particle positions were then linked between frames into particle trajectories, which were subsequently used to compute the local strain and strain correlations.

\section{Mode Coupling Theory}

Strain correlations  in  fluid states can be  considered  based on a connection to velocity fluctuations \cite{Illing2016}.  The pertinent quantities and predictions shall be summarized here as basis for the later tests in experiment and simulation.

  The  average correlation of accumulated strains at positions separated by distance $\bf r$  is defined as
\begin{equation}\label{mct1}
{ C}_{\boldsymbol{\varepsilon}}({\bf r},t) =   \langle
\boldsymbol{\varepsilon}({\bf r}+{\bf r}_0,t) \; \boldsymbol{\varepsilon}({\bf r}_0,t)   \rangle\;   ,
\end{equation}
where  the strain field 
accumulated over time $t$ follows from the  integral of the symmetrized velocity gradient tensor:
\begin{equation}\label{mct2}
\boldsymbol{\varepsilon}({\bf r},t)
=  \int_{t_0}^{t_0+t} dt' \; \frac 12 \left( \nabla  {\bf v}({\bf r},t')+ (\nabla  {\bf v}({\bf r},t'))^\top \right)\; ,
\end{equation}
where $\top$ here stands for transpose operator. This relation rests on the identification of the velocity field as time-derivative of the displacement field, $\dot{\bf u}_{\bf q}={\bf v}_{\bf q}$, and the  familiar (linearized) relation between displacement and strain field.
In an equilibrated and homogeneous system, the correlation ${C}_{\boldsymbol{\varepsilon}}$  does not depend on the arbitrarily chosen ${\bf r}_0$ and $t_0$, which will be used in the averaging performed in the simulations.  To compute ${C}_{\boldsymbol{\varepsilon}}$ and simplify the analysis, we apply a Fourier-Laplace transformation according to ${C}_{\boldsymbol\varepsilon}({\bf q},s)= \int_0^\infty dt \int d{\bf r}
e^{i{\bf q\cdot r}-st}  { C}_{\boldsymbol\varepsilon}({\bf r},t)$. The advantage is that the transform ${C}_{\boldsymbol\varepsilon}({\bf q},s)$ is directly related to the autocorrelation tensor of the velocity field, which is related to  memory kernels in the Zwanzig-Mori approach. This approach uses the Fourier-Laplace space to perform the long-wavelength analysis of (generalized) hydrodynamics, which will be our basis for establishing the strain correlations in the far-field \cite{Hansen1990}.  The Zwanzig-Mori decomposition applied to ${\rm C}_{\boldsymbol{\varepsilon}}({\bf q},s)$  gives the following result for a shear element \cite{Illing2016}:
\begin{eqnarray}\label{mct3}
 { C}_{\varepsilon_{xz}}({\bf q},s)  &=&     \left( \frac{q_x^2+ q_z^2}{4} - \frac{q_x^2q_z^2}{q^2} \right) \frac{2}{s^2} { K}^{\perp}_q(s)\\\notag
 &+& \left( \frac{q_x^2q_z^2}{q^2} \right) \frac{2}{s^2} { K}^{\|}_q(s)
 \end{eqnarray}
 Here ${\bf K}=\langle {\bf v}_{\bf q}^*(t){\bf v}_{\bf q}\rangle$ is the auto-correlation tensor of the velocity field, and superscripts indicate its two parts, longitudinal ($ ^\|$) and transverse ($ ^\perp$).
The auto-correlation functions of the  velocities are connected to memory kernels ${\bf G}$ via
\begin{equation}\label{mct4}
{ K}^{(i)}_q(s) = \frac{v_{\rm th}^2}{s + \frac{q^2}{\rho} { G}^{(i)}_q(s)}\;,
\end{equation}
 which again arise from  longitudinal ($^{(i)}$$\to$$ ^\|$) or transverse ($^{(i)}$$\to$$ ^\perp$) fluctuations \cite{Hansen1990}.  Here $\rho$ denotes mass-density and  $v_{\rm th}=\sqrt{k_BT/m}$ the (one-dimensional)  thermal velocity of particles with mass $m$.  The fluctuating  force memory kernels ${ G}^{(i)}_q(s)$ generalize the shear and longitudinal viscosity to finite frequencies and wavevectors.
It is the law of momentum conservation which causes the appearance of hydrodynamic poles in ${\bf K}$ in Eq.~\eqref{mct4}. They capture transverse momentum diffusion and longitudinal compressional waves in the long-wavelength and low-frequency limit of standard hydrodynamics, where the kernels can be replaced by transport coefficients (viz.~viscosities).  The general formal expressions of the memory kernels in the Zwanzig-Mori approach are
\begin{eqnarray}\label{mct5}
{\bf G}({\bf q},t) &=&    \left[\frac{{\bf q}{\bf q}}{q^2 \kappa^T_q} +   \frac{\rho}{(mv_{\rm th}q)^2}\,\langle {\bf F}_{\bf q}^*(t_{\cal Q})  {\bf F}_{\bf q} \rangle  \right]
  \notag \\
 &=& \frac{{\bf q}{\bf q}}{q^2}\; {\rm G}^\|_q(t)  + ({\bf 1} - \frac{{\bf q}{\bf q}}{q^2})\; {\rm G}^\perp_q(t)\;
\end{eqnarray}
where $\kappa^T_q$ denotes the generalized $q$-dependent isothermal compressibility, and the subscript at $t_{\cal Q}$ indicates that the time evolution of the memory kernel follows the reduced dynamics devoid of hydrodynamic poles.  The force fluctuations arise in the law of momentum  conservation, $\partial_t m{\bf v}_{\bf q}(t) = {\bf F}_{\bf q}(t)$. Because of ${\bf F}_{\bf q}\to0$ for $q\to0$, they can be reformulated in terms of stress tensor elements \cite{Hansen1990}.

The consequences of the obtained relations for strain correlations in supercooled liquids can be discussed within a simple model. In order to encode the growth of the viscosity when approaching the glass transition, a simple ansatz following Maxwell can be made for the transversal memory kernel \cite{Maier2017}: ${\rm G}^\perp_q(t) \approx G^M(t)=G^\bot_\infty e^{-t/\tau}$ which should be valid  for small wavevectors. Here $G^\bot_\infty$ is the macroscopic static shear modulus measurable in rheological spectra at intermediate frequencies $1/\tau\ll \omega\ll 1/\tau_0$ (with $\tau_0$ some microscopic time scale), and $\tau$ is Maxwell's relaxation time. The growth of $\tau$ explains the increase of the viscosity, $\eta=G^\bot_\infty \tau$. Considering the system as incompressible and thus neglecting the longitudinal contribution to Eq.~\eqref{mct3} for the time being, one notices that the strain correlations obey a scaling law (in $d$ dimensions):
\begin{equation}  \label{mct6}
{ C}_{{\varepsilon}_{xz}}({\bf r},t) = \frac{1}{\xi^{d}}(\frac{v_{\rm th}}{\vs})^2\tilde{ C}_{\varepsilon_{xz}}({\bf r}/ \xi,t / \tau), \mbox{ where } \xi=\vs \tau\,,
\end{equation}
with an universal function \(\tilde{ C}_{{\varepsilon}_{xz}}(\tilde{\bf r},\tilde{t})\).
Here the  velocity \(\vs = \sqrt{{G^\bot_\infty}/{\rho}}\) of (high-frequency) transverse sound was introduced.
Anticipating that the length \(\xi\) becomes large in the considered supercooled fluids due to rapid growth of the relaxation time \(\tau\), the limit of \(\xi \to\infty\) can be taken. It immediately predicts the appearance of spatial power-law correlations in the far-field,
\({C}_{\boldsymbol{\varepsilon}_{xz}}\propto r^{-d}\). To be precise, for distances large compared to microscopic lengths but shorter than the size of the elastic domain, \(a \ll r \ll \xi\), the transverse part reads
 \begin{equation}\label{mct7}
\tilde{C}^{\bot}_{{xz}}(\tilde{\bf r},\tilde{t}) = \frac{3}{8\pi}\, \frac{\tilde{J}^M(\tilde t)}{\tilde r^3} \;  \frac{\tilde r^2(\tilde x^2+\tilde z^2)-10\tilde x^2\tilde z^2}{\tilde r^4}.
\end{equation}
The spatial dependence follows Eshelby's far-field pattern of strain in an isotropic elastic solid (e.g.~in the plane $y=0$, a four-fold angular pattern results from the dependence on $x^2z^2$) \cite{picard2004}. The long-range pattern is built-up by the diffusion of transverse momentum, which enters a factor $K^\perp\propto 1/(qL)^2$ in Eq.~\eqref{mct4} for small frequencies.  Yet, because of the bigness of $\tau$, the temporal evolution still contains the fluid limit additionally to the expected solid limit. Both are contained in the (scaled) creep compliance $\tilde J$, which is the solution of $\int_0^{\tilde{t}} \tilde{G}(\tilde{t}-\tilde{t}') \tilde{J}(\tilde{t}')d\tilde{t}' = \tilde{t}$, where $\tilde{G}=G^\perp_0/G^{\bot}_\infty$, in the general case. The one  corresponding to Maxwell's ansatz for the modulus reads:
$\tilde{J}^M(t/\tau) =  1 + \frac{t}{\tau}$.
For times short compared to the relaxation time, the pattern is constant as expected for a solid. Only in fluid states, where $\tau$ is finite, also the limit $t\gg\tau$ can be accessed, where the strain  keeps its spatial correlations but grows linearly with time. The overall prefactor then contains the viscosity. MCT calculations recover both limits and provide quantitative predictions e.g.~for $\tau$ and $G^{\bot}_\infty$ \cite{Illing2016}.

In compressible systems, also longitudinal velocity fluctuations contribute to the shear-strain in Eq.~\eqref{mct3}.
The correlation function then can be written as the sum of transverse and longitudinal contributions $\tilde{ C}_{{xz}}(\tilde{\bf r},\tilde{t}) = \tilde{C}^{\bot}_{{xz}}(\tilde{\bf r},\tilde{t})  + \tilde{C}^{\|}_{{xz}}(\tilde{\bf r},\tilde{t})$, where  the contribution of longitudinal velocity correlations to the far field reads:
\begin{eqnarray}\label{mct8}
\tilde{C}_{{xz}}^{\|}(\tilde{\bf r},\tilde{t}) &\to& \frac{3}{8\pi}\, \frac{g^\perp\, \tilde{J}^\|(\tilde{t})}{\tilde r^3} \;  \frac{10\tilde x^2\tilde z^2}{\tilde r^4}
  ,\quad \mbox{for } \xi\to\infty.
\end{eqnarray}
Here, $g^\perp$  compares the frozen-in (macroscopic, viz.~${\bf q}=0$) shear modulus,  ${G}^\bot_{\infty}={ G}^\perp_{0}(t\to \infty)$, to the isothermal compressibility, viz.~$g^\perp=\kappa^T_0\, {G}^\bot_{\infty}$. This combination arises because the longitudinal velocity correlations are bounded by the fluid compressibility $\kappa^T_q$
and because of the factor $(v_{\rm th}/\vs)^2$, which is chosen as prefactor in the scaling law Eq.~\eqref{mct6}; it  contains $1/G^\bot_\infty$.

The longitudinal compliance function $\tilde{J}^\|$ takes an explicit form when entering again a Maxwell ansatz for the longitudinal kernel with, for simplicity, the same relaxation time $\tau$ as in the transverse modes. The function  $\tilde{J}^\|$  is not universal but depends on parameter $g^\|$  which compares the frozen-in, macroscopic longitudinal modulus, $G^\|_\infty=G^\|_{0}(t\to\infty)$, to the isothermal compressibility, viz.~$g^\|=\kappa^T_0\,G^\|_\infty$. The result in Laplace space reads:
$\tilde{J}^\|(\tilde{s})= \tau(1+\tilde{s})/(\tilde{s} (1+\tilde{s}) + (g^\|-1) \tilde{s}^2)$.  It only contributes appreciably to the strain pattern of the solid, where   $\tilde{J}^\|(\tilde{t}\to0)=1/g^\|$, while it becomes negligible in the fluid state, where it remains bounded as well, $\tilde{J}^\|(\tilde{t}\to\infty)=1$. Even in glass, the correction to Eq.~\eqref{mct7} will be minor because the value of $g^\perp/g^\|=G^\bot_\infty/G^\|_\infty$ can be expected to be small.


\section{Computer Simulations}

For the MD part of this study, we use the well known Kob-Andersen binary (80:20) Lennard-Jones mixture \cite{Kob1994} at a total density of $\rho\myeq \rhoA+\rhoB \myeq 1.2$. At this number density and composition,
the mode coupling critical temperature (a measure of the proximity to the glass
transition) of the model is $\Tc \myeq 0.435$ \cite{Kob1995a}. A and B particles
in the model interact via $U_{\mathrm{LJ}}(r)\myeq
4\epsilon_{\alpha\beta}[(d_{\alpha\beta}/r)^{12}-(d_{\alpha\beta}/r)^6],$ with
$\alpha,\beta\myeq {\mathrm{A,B}}$, $\epsilon_{\mathrm{AB}}\myeq
1.5\epsilon_{\mathrm{AA}}$, $\epsilon_{\mathrm{BB}}\myeq
0.5\epsilon_{\mathrm{AA}}$, $d_{\mathrm{AB}}\myeq 0.8d_{\mathrm{AA}}$,
$d_{\mathrm{BB}}\myeq 0.88d_{\mathrm{AA}}$ and $m_{\mathrm{B}}\myeq
m_{\mathrm{A}}$. In order to enhance computational efficiency, the potential is
truncated at twice the minimum position of the LJ potential, $\rcab\myeq 2.245
d_{\alpha\beta}$. The parameters $\epsilon_{\mathrm{AA}}$, $d_{\mathrm{AA}}$ and $m_{\mathrm{A}}$ define the units of energy, length and mass, respectively.  The unit of time is a combination of these units, $\tauLJ \myeq d_{\mathrm{AA}}\sqrt{m_{\mathrm{A}} / \epsilon_{\mathrm{AA}}}$. Unless otherwise stated, the simulation box is a cube of length $L=100$, containing $1.2\times10^6$ particles. All the simulations reported here are performed using
LAMMPS~\cite{Plimpton1995} with a discrete time step of $dt \myeq 0.005$. The model has been investigated in previous works, addressing various issues such as non-Newtonian rheology \cite{Berthier2002a,Varnik2006d}, heterogeneous plastic deformation and flow \cite{Varnik2003,Hassani2016} and structural relaxation under shear \cite{Berthier2002a,Varnik2006b}. It is noteworthy that, even though a crystalline equilibrium state does exist for this model~\cite{Pedersen2018,Ingebrigtsen2018}, the time necessary to reach this state exceeds by orders of magnitude the simulation times relevant for the present study. More specifically, we have carefully scrutinized this issue and observe no signature of crystallization at all the simulations whose results are reported here.

\section{Strain fluctuations and correlations thereof}
Strain fluctuations are evaluated via the following procedure. First, using particle positions, $\vec{r}_i$ ($i$ is the particle index), at times $t_0$ and $t_0+t$, displacement vectors are defined: $\vec{u}_i(t)=\vec{r}_i(t_0+t)-\vec{r}_i(t_0)$. In order to reduce the statistical noise, the thus obtained displacement field is averaged over a length scale, $w$, usually of the order of the nearest neighbor distance. This coarse-graining process is performed via $\vec{u}^\text{CG}(\vec{r},t)={\sum_{i}{\vec{u}}(\vec r_i, t)\phi(||\vec r-\vec r_i||)}/{\sum_{j}\phi(||\vec r-\vec r_j||)}$, using the coarse-graining function, $\phi(r)=\frac{1}{(\pi w^2)^{3/2}}e^{-(r/w)^2}$ \cite{Goldenberg2007,Hassani2018}. The sum is performed for all the particles within a sphere of radius $\sim w$ around point $\vec r$. The strain tensor is obtained as $\boldsymbol\varepsilon(\vec{r},t)=(\nabla \vec{u}^\text{CG}(\vec{r},t)+(\nabla \vec{u}^\text{CG}(\vec{r},t))^\top)/2$.  It is important to emphasize that, due to the absence of shear in the present study, $\varepsilon$ is identical to the quiescent  (coarse-grained)  strain fluctuation.

Taking the specific example of $xz$-component of the strain tensor, spatio-temporal correlations of shear strain fluctuations are defined as $\Cexz(\vec{r}_{||},t)= \langle \exzna(\vec{r}_{||} + \vec{r}_0, t)  \exzna(\vec{r}_0, t)\rangle$; see Eq.~\eqref{mct1}. The use of $\vec r_\parallel$ recalls that the evaluation is performed within thin layers (slabs) parallel to the $xz$-plane. In the absence of externally imposed shear, isotropy implies that all non-diagonal components of the strain tensor are equivalent and that the corresponding correlation functions provide exactly the same information. We have explicitly checked this property for $\Cexz$, $\Cexy$ and $\Ceyz$, evaluated within the corresponding $xz$, $xy$ and $yz$ planes, respectively  (data not shown). For a quantitative comparisons with theoretical predictions, it is useful to integrate the correlation function with respect to the polar angle and only keep the dependence on radial distance and time:
\begin{eqnarray}\label{eq:C44}
C^4_4(r, t) &=& \frac{1}{\pi}\int_0^{2\pi}{\Cexz(\vec{r}_{||},t)\cos(4\theta}) \diff \theta,\\
&=& \frac{\Cs(t)}{r^3}. \label{eq:C44-Cs}
\end{eqnarray}
where \(\theta\) is the polar angle and \(\vec r_\parallel=r_\parallel (\cos(\theta), 0, \sin(\theta))\) within the $xz$-plane at \( y=0 \). In writing the second line above, Eq.~(\ref{eq:C44-Cs}), we have made use of the \(1/r^3\)-scaling behavior of the strain-strain correlation function for \(a\ll r\ll \xi\). A straight-forward asymptotic analysis of \(\tilde{J}_M(\tilde t)\) and \(J^\|(\tilde t)\) appearing in Eqs.~(\ref{mct7}) and (\ref{mct8}), respectively, leads to the following prediction for the correlation amplitude,
\begin{equation}
C_s(t)= \begin{cases}
\dfrac{15\rho \kB T}{32 \pi m}(\dfrac{1}{G^\bot_\infty}-\dfrac{1}{G^\|_\infty})~~~\text{glass}\\
\dfrac{15\rho \kB T}{32 \pi m}\dfrac{t}{\eta}~~~\text{supercooled~liquid}
\end{cases}
\label{eq:Cs}
\end{equation}
Here, the longitudinal contribution in the compressible glass is included.

The same strain correlation analysis is performed on experimental and simulational data, taking the particle trajectories, determined over eight time points in the former case, as input.

Before presenting the results, we remark that the use of a coarse graining procedure does not bias spatial correlations of strain but only reduces the statistical noise (data not shown). This applies both to the experimental as well as simulated results.

\section{Results}
\begin{figure}
	\begin{center}
(a)\includegraphics[width=8cm]{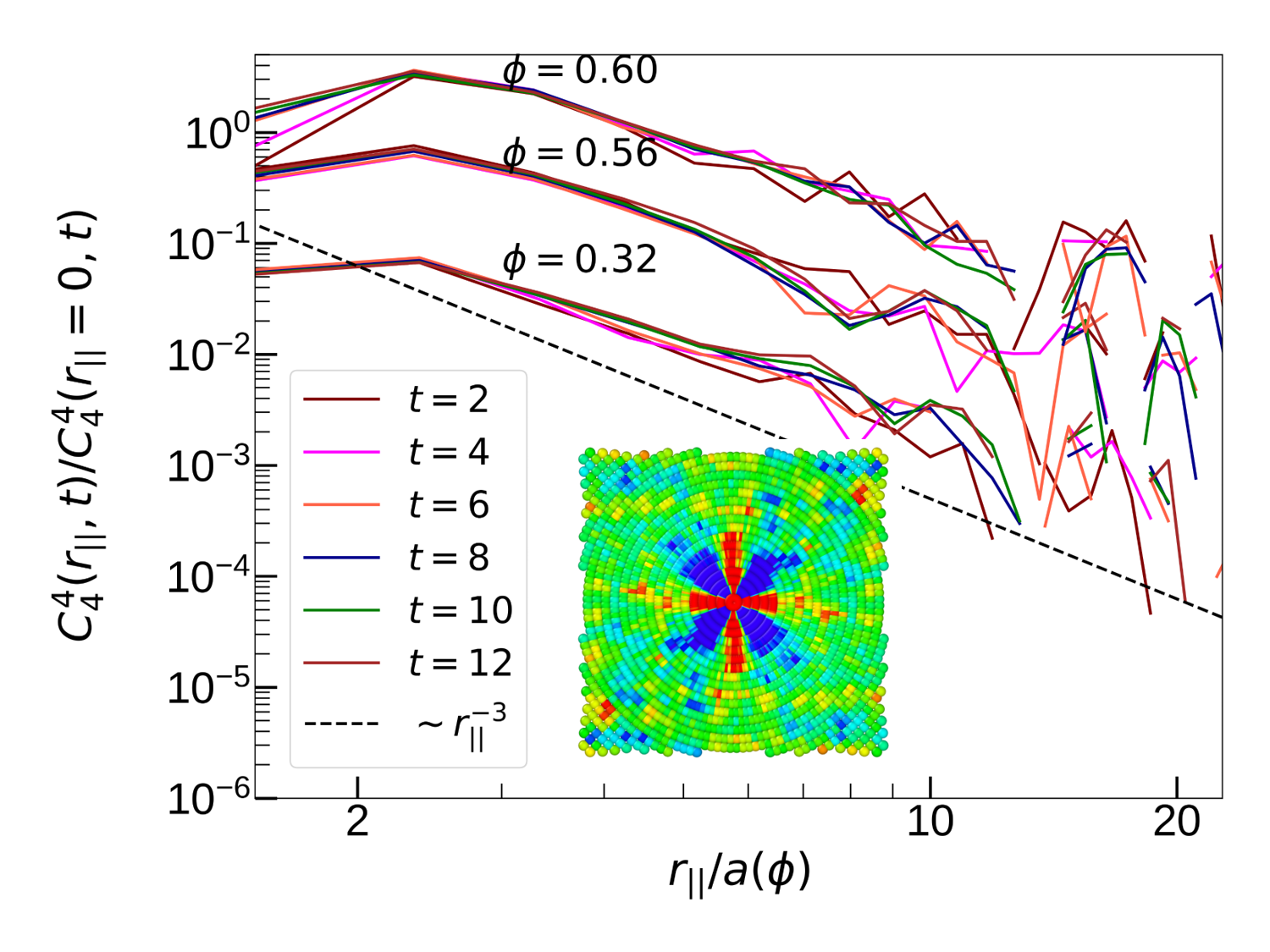}\\
(b)\includegraphics[width=8cm]{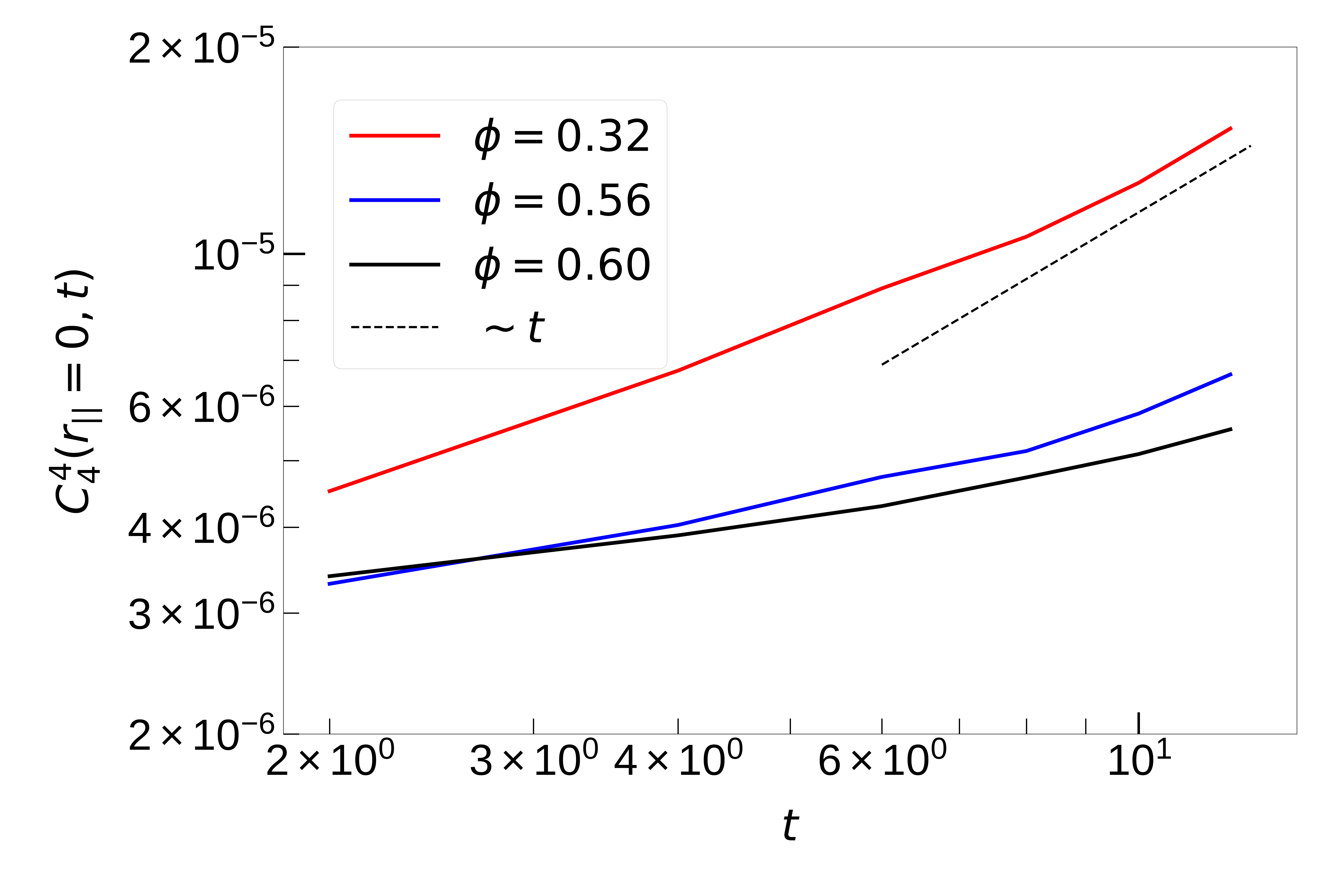}
		\caption{Experimental results on normalized correlations of strain fluctuations in a hard sphere colloidal glass former. Panel (a) shows the correlation function versus in-plane-distance, $r_\parallel/a(\phi)$, after integration over the polar angle, Eq.~(\ref{eq:C44}).  Here, $a(\phi)$ is the average inter-particle distance, obtained from the first minimum of the radial pair distribution function. The data are presented for three volume fraction of $ \phi = 0.32$, $0.56$ and $0.60$. Strain correlations are computed over time intervals of $t = 2, 6, 8, 10$ and $12$min. For all the three volume fractions, the data show approximately a power-law decay with an exponent of $-3$. In (b), amplitude of the correlations for the three volume fraction is shown. It exhibits a quasi-plateau close to and at the glassy state ($\phi=0.56$ and $\phi=0.6$), while it grows linearly in the supercooled regime ($\phi=0.32$). The inset in (a) shows color scale plot of the correlation function for $\phi=0.32$, highlighting the four-fold symmetry.}
		\label{fig:Exp}
\end{center}
\end{figure}

We observe that strain correlations in the quiescent colloidal glass exhibit quadrupolar symmetry, as shown in the inset of Fig.~\ref{fig:Exp}a. This observation is in qualitative agreement with previous two-dimensional simulations~\cite{Chattoraj2013} and experiments~\cite{Jensen2014}. To investigate these quadrupolar correlations in more detail, we project them onto the corresponding circular harmonic given by $\cos(4\theta)$, Eq.~(\ref{eq:C44}), and study their radial decay. The resulting projected correlation functions are shown as a function of distance in Fig.~\ref{fig:Exp}a.  Remarkably, all correlation functions decay with the same power of $-3$, irrespective of the volume fraction and the time interval, over which the strains are computed. While the power-law decay is thus robust over the volume fractions from fluid to dense glass, the correlation amplitude varies, showing different trends for fluid and glass. This is shown in Fig.~\ref{fig:Exp}b, where we plot the correlation magnitude at the origin ($r=0$) as a function of observation time interval for the three volume fractions. With increasing time $t$, the lowest volume fraction shows a growth of correlation amplitude, approaching a power of $\sim 1$ towards the end; thus, the correlation amplitude grows with the amount of strain accumulated during the observation time interval. In contrast, the higher volume fractions show only a shallow increase of correlations, indicating a quasi plateau at small time intervals. These trends, which agree qualitatively with the predictions in Eq.~\eqref{eq:Cs}, are confirmed in other experimental data sets of quiescent colloidal glasses over similar ranges of volume fraction (data not shown).

\begin{figure}
\begin{center}
\includegraphics[width=8cm]{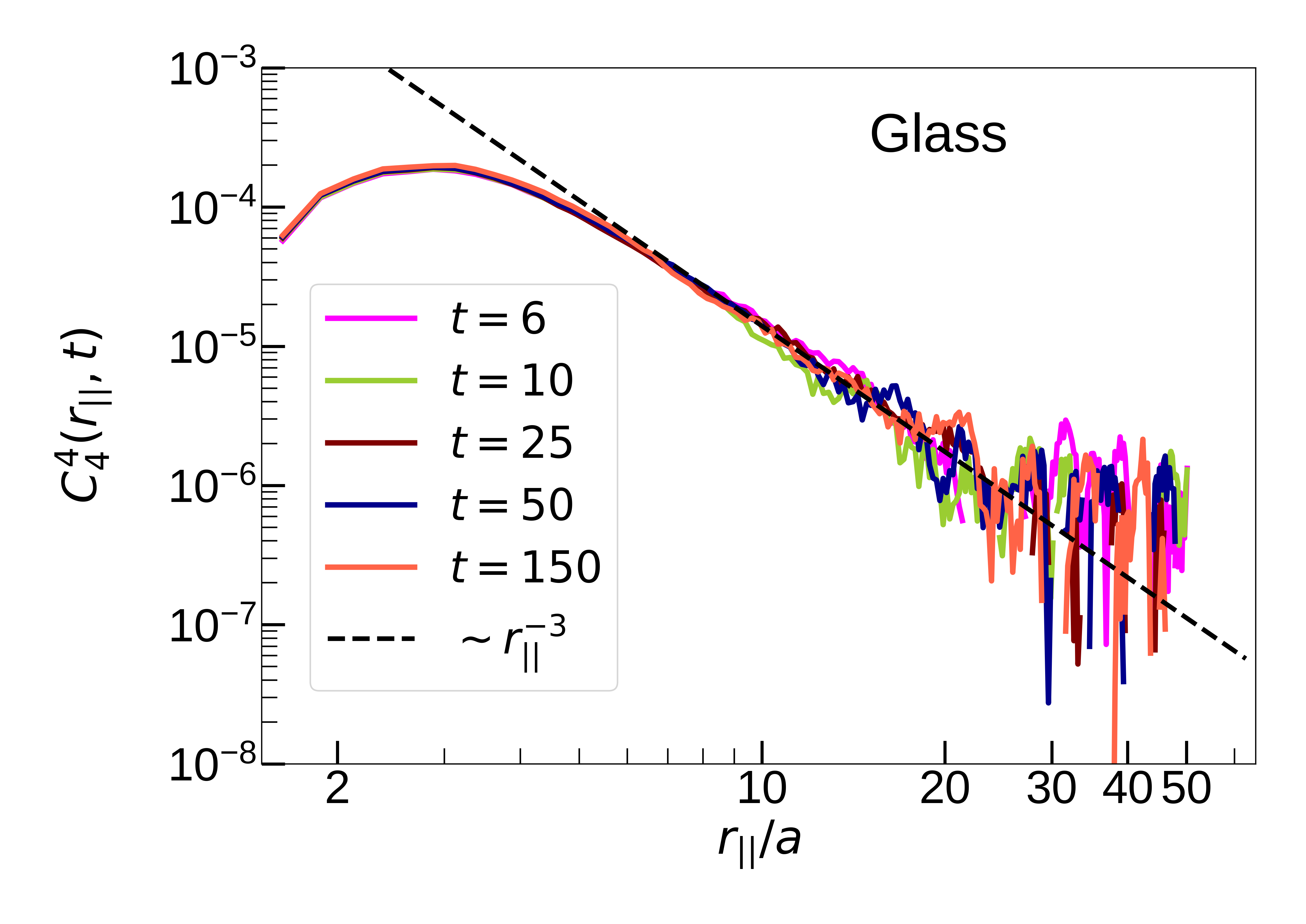}
\end{center}
\caption{Normalized correlations of strain fluctuations obtained from MD simulations of a generic binary LJ glass former at a temperature of $T=0.2$ (glassy state). Each curve corresponds to a time, $t$, used to evaluate the strain fluctuation. At distances large compared to a particle diameter, the correlation function decays with a power-law with an exponent of $-3$ for all the time intervals investigated. The correlation amplitude is shown in Fig.~\ref{fig:corr-ampl-vs-t}.}
\label{fig:MD-glass}
\end{figure}

\begin{figure}
	\begin{center}
		\includegraphics[width=8cm]{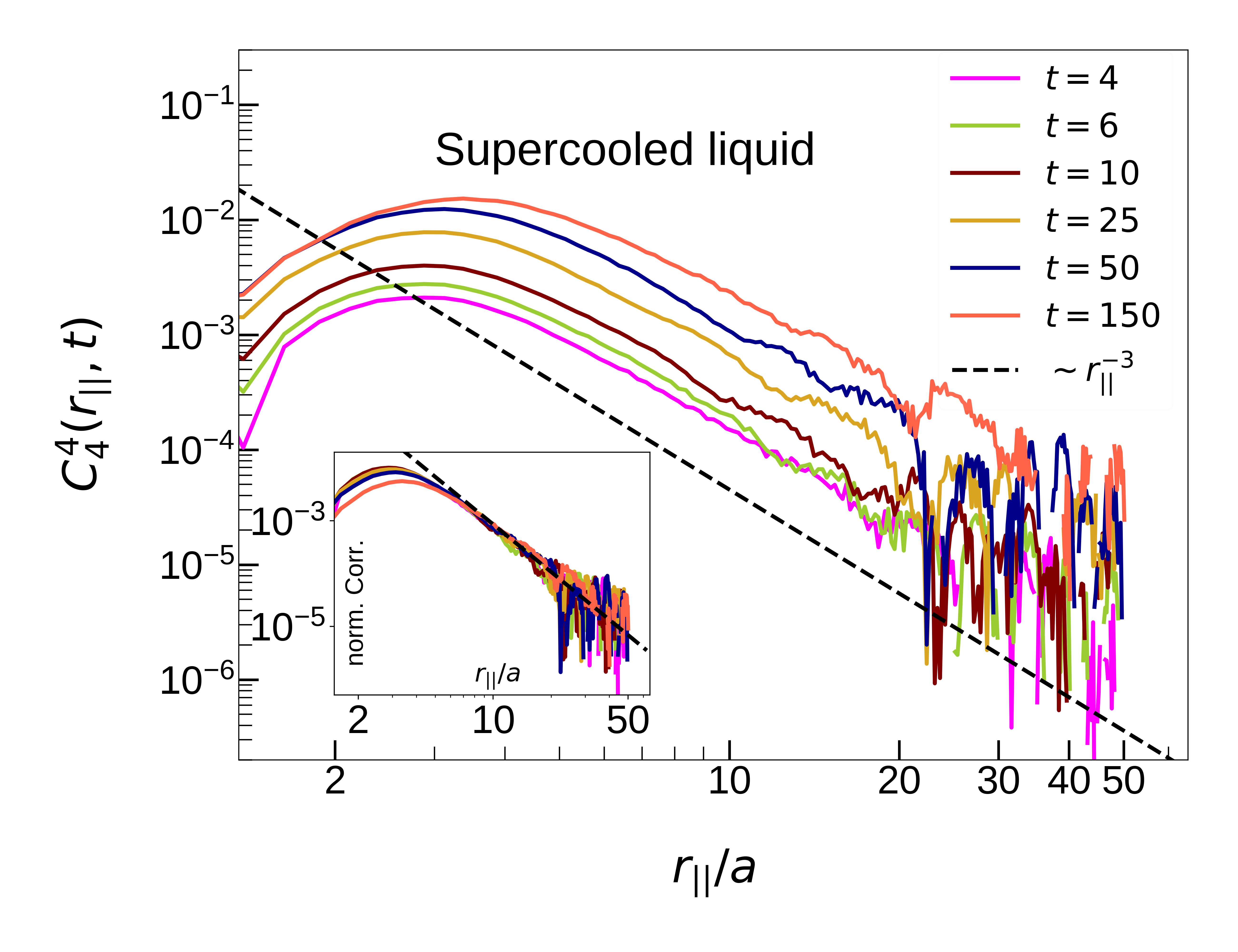}
	\end{center}
	\caption{The same quantities as in Fig.~\ref{fig:MD-glass} but at a temperature of $T=0.7$ (supercooled liquid state).	In quite similarity to the glassy state, a power-law decay of $\sim r^{-3}_\parallel$ governs the behavior of correlations at sufficiently large distances. However, in contrast to the glassy state, the liquid character of the system manifests itself in a growth of the correlation amplitude with time. The correlation amplitude is shown in Fig.~\ref{fig:corr-ampl-vs-t}.}
	\label{fig:MD-supercooled}
\end{figure}

To better quantify these trends and elucidate their origin, we study strain correlations in MD simulations of three-dimensional glasses using the binary LJ model described above. The resulting correlation function of shear strain fluctuations is shown in Fig.~\ref{fig:MD-glass} for a temperature of $T=0.2$ ($\Tg \approx 0.4$~\cite{Varnik2006d}), which belongs to the glassy state of the model. For all the times $t$ used to evaluate the strain, the (non-normalized!) correlation functions obey a master curve with a power-law decay, $1/r^3$, for distances large compared to the particle size, similar to the experiments. This is the domain relevant for continuum mechanics, where effects arising from molecular scale structure become irrelevant. It is interesting to compare this behavior of a glass to that of a supercooled liquid, Fig.~\ref{fig:MD-supercooled}. In this case, the strain correlation still obeys a power-law with exponent -3 but the amplitude of correlations increases with time.

\begin{figure}
	\begin{center}
		\includegraphics[width=8cm]{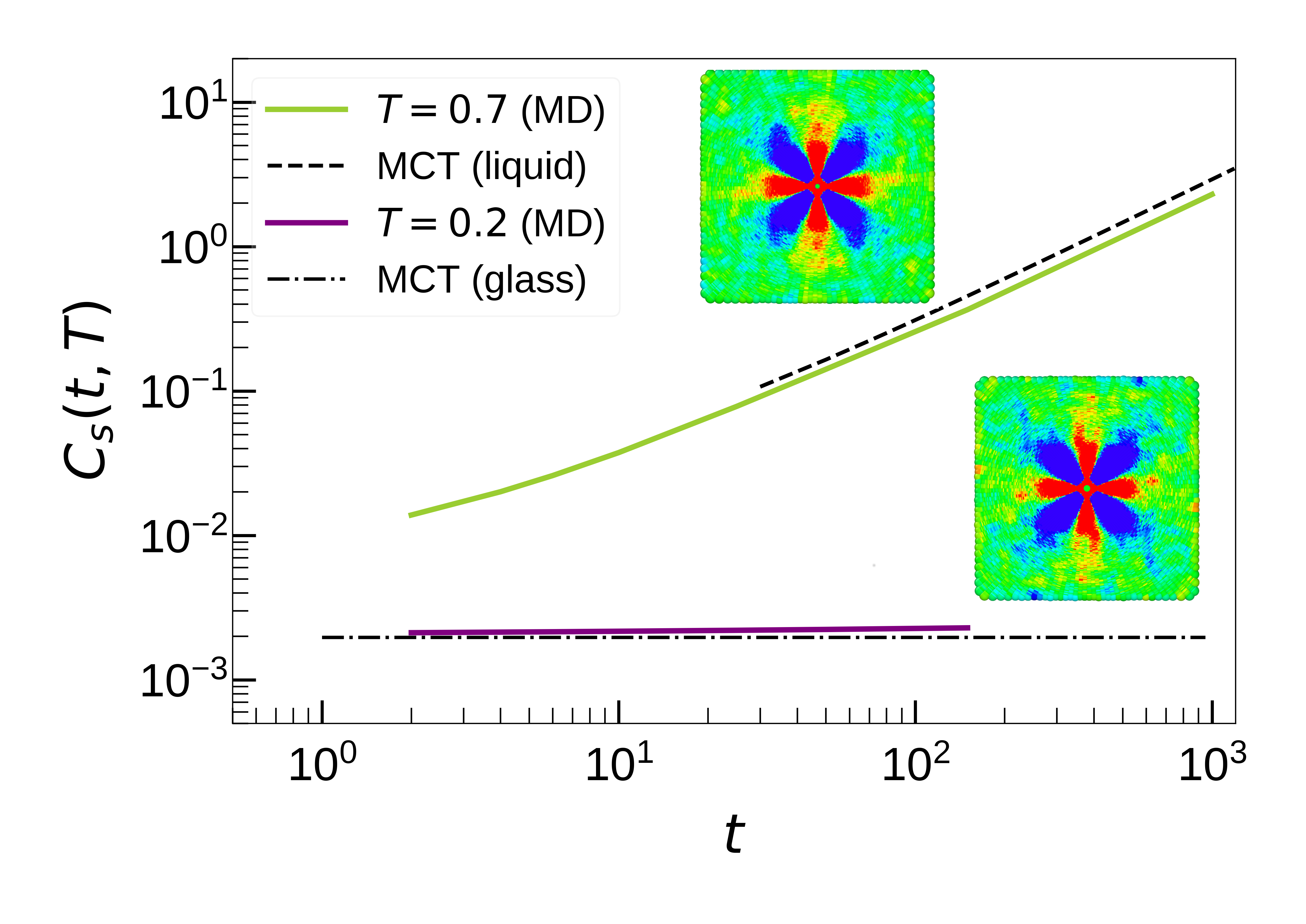}
	\end{center}
	\caption{The correlation amplitude versus time at temperatures of $T=0.7$ (supercooled liquid state) and $T=0.2$ (glass). The dashed lines give theoretical predictions for the respective cases using Eq.~(\ref{eq:Cs}) ($\rho=1.2$, $\kB=1$, $G^\bot_\infty(T=0.2)=15$~\cite{Varnik2004}, $G^\|_\infty(T=0.2)=86.3$, $\eta(T=0.7)=42$). The color scale plots show the correlation function at respective temperatures, highlighting the four-fold symmetry both in the glassy state and in the supercooled liquid.}
	\label{fig:corr-ampl-vs-t}
\end{figure}

A quantitative analysis of the correlation amplitude, $\Cs(t)$, is depicted in Fig.~\ref{fig:corr-ampl-vs-t}. It is noteworthy that no fit parameter is used for this comparison. Rather, all the constants entering theoretical predictions are evaluated from independent simulations. Simulation results agree well with theoretical predictions both in the glass and in the supercooled state.

\section{Size of elastic domains in the liquid state}

The length-scale associated with a solid-like response in the supercooled state is predicted to be given by  \(\xi \sim \tau \sqrt{G^{\bot}_\infty/\rho}\), where \(\tau\) is a structural relaxation time, and \(G^{\bot}_\infty\) the high frequency shear modulus. Recalling that the speed of sound is given by \(\vs=\sqrt{G^{\bot}_\infty/\rho}\), it is seen that \(\xi \sim \tau \vs\)~\cite{Maier2017}. Using this relation, the four-fold pattern and the characteristic \(1/r^3\) decay of spatial correlations finds a simple interpretation within MCT. A local deformation event generates a signal which propagates with the speed of sound in the surrounding medium. As long as no structural relaxation takes place during the propagation of the signal, the region "visited" by the signal appears as an elastic medium. This ceases to be the case as time exceeds the structural relaxation time.
It is noteworthy that the idea of a 'solid-like' region in a liquid has been around for a while. 
Dyre, for example, predicted that the size, \(l\), of such a region scales as \( l\sim (c_\text{glass} \tau)^{1/4}\), where \(c_\text{glass}\) is the longitudinal sound velocity in the glass and \(\tau\) is the time between two 'flow events' in a sphere of radius \(l\)~\cite{Dyre1999}. However, transverse collective hydrodynamic modes, which preserve momentum and which give rise to long range correlations have not been considered there.

A correlation length, which is closely tied to---and linearly grows with---the structural relaxation time has also been reported in the case of a four-point dynamic structure factor, constructed from the full complex self intermediate scattering function~\cite{Flenner2015,Flenner2015b}. Interestingly, the increase of the correlation length is attributed to the growth of transient elastic response. A growing length has also been observed in the related cross-over from diffusion to wave propagation in the transverse momentum correlations~\cite{Torchinsky2012}.

In a typical liquid, the speed of sound is of the order of \(1000\text{m/s}\). Using \(\tau \sim 10\text{ps}=10^{-11}\text{s}\) for the structural relaxation time, one thus obtains \(\xi \sim 10^{-8}\text{m}=10\text{nm}\), corresponding already to 10-50 molecular diameters. In the glassy state, \(\tau\sim 100\)s, which yields \(\xi\sim 1000\)m, a truly macroscopic length scale. To estimate \(\xi\) for our simulations, we have performed a thorough analysis of the stress autocorrelation function in the quiescent system and the frequency-dependent elastic modulus under oscillatory shear. As a result of these investigations whose details will be reported elsewhere, we have determined \(G^{\bot}_\infty\) and \(\tau\) for the present binary LJ model both in the supercooled state and in the glass. In the supercooled state (\(T=0.7\)), we find \(G^{\bot}_\infty \approx 13\) and \(\tau \approx 3\), which gives \(\xi \approx 10\). Interestingly, this estimate is quite close to the one obtained from the Maxwell approximation for the stress relaxation time, $\tau=\eta/G^\bot_\infty$, which leads to \(\xi=\eta/\sqrt{\rho G^\bot_\infty}\) and thus \(\xi\approx 11\), where \(\eta(T=0.7)=42\) was used. In the glassy state (\(T=0.2\)), the shear modulus raises only slightly (\(G^{\bot}_\infty \approx 16\)) but the relaxation time grows by orders of magnitude, \(\tau\ge 10^5\), leading to \(\xi(T=0.2) \ge 10^5\), far beyond the simulation box size.

The fact that, depending on temperature, the size of the elastic domain in the supercooled state can be of the order ten particle diameters is encouraging to study a possible cross-over from the long-range Eshelby-like quadrupolar correlations to a different behavior, characteristic of the liquid state. The present set of data, however, does not show a clear signature of such a cross-over. We shall here recall that the above estimate of the length scale $\xi=\tau\vs$ is based on a scaling argument and thus contains an a priori unknown numerical factor. Future studies with larger simulation box sizes could help to elucidate this issue.

To estimate $\xi$ for the experiments, we have to take a different route as the description of the overdamped colloidal system differs in one respect.  Here, forces on the particles arise also from the solvent. These forces can be considered rapid and fluctuating, but violate momentum conservation because the solvent on average exerts friction. Modeling the colloidal particles by Langevin equations \cite{Hess1983}, the memory kernels in Eq.~\eqref{mct4} are replaced by $q^2 G^{(i)}_q(s)/\rho  \to \frac{\zeta}{m}+ q^2 G^{(i)}_q(s)/\rho$. Here, $\zeta$ can (for simplicity) be taken as Stokes friction coefficient. Considering the overdamped limit, where friction dominates over inertia, $\zeta\gg ms$, the strain fields can be estimated for colloidal dispersions. The strain patterns described in Eqs.~\eqref{mct7} and \eqref{mct8} keep their form, yet the expression for the correlation length changes to
$\xi^2=G^{\bot}_\infty/(n\zeta_0) \tau$ \cite{Illing2016}, where $n$ is the particle density. Apparently, signals from deformation events now propagate by a random walk with diffusion coefficient given by elastic relative to viscous forces.  We estimate the corresponding correlation length $\xi$ from measurement of the shear modulus taken from~\cite{vanderVaart2013}, and relaxation time taken from~\cite{vanMegen1998}. In the volume-fraction regime from $\phi = 0.32$ to $0.6$ investigated here, the shear modulus changes from some $10 k_BT/R^3$ to $\sim 200 k_BT/R^3$, while the glass relaxation time changes from $\sim 10 \tau_0$ to larger than $10^6 \tau_0$, where $\tau_0$ is the relaxation time at infinite dilution defined by $\tau_0 = R^2 / (6D_0)$ with the diffusion coefficient $D_0\zeta_0 = k_BT$. Using these values, one finds that the length $\xi$ increases from $\xi/R \sim 36$ at $\phi = 0.32$ to $\xi/R \sim 2500$ at $\phi = 0.56$ and to $\xi/R \gtrsim 37000$ at $\phi = 0.6$. This indicates that also in the experiment, the crossover to the fluid regime may be outside the accessible length scale range, even at the lowest volume fraction studied ($\phi = 0.32$). Indeed the strain correlations in Fig.~\ref{fig:Exp}a show an $r^{-3}$ decay over the full range, without clear signature of a finite correlation length.

\section{Conclusion}
In this work, we investigated correlations of strain fluctuations in quiescent glasses and supercooled liquids via experiments and computer simulations in three dimensions. Both in the glassy state and in the supercooled liquid, experimental and simulated strain correlations decay with a $1/r^3$ power-law decay, reminiscent of  Eshelby's pattern. The four-fold symmetry is also preserved in the both cases investigated. The spatial pattern arises from diffusive transport of transverse momentum coupled into the strain field. The difference between the glassy and liquid states manifests itself in the time dependence of the correlation amplitude which forms a plateau in the glassy state whereas it grows linearly in the supercooled liquid state. Qualitatively, the strain amplitude follows a Maxwellian compliance in response to thermal stresses. All these observations are rationalized within the recently developed MCT, applied here to the present 3D problem. The length scale associated with the size of solid-like response in the supercooled regime is given by the distance traveled by sound during the structural relaxation time. In overdamped systems, it is the distance elastic forces diffuse in time $\tau$. It ranges from tens of nanometer in the high temperature normal liquid state to macroscopic lengths of the order of kilometer in the glassy state. From this follows that accompanying the dramatic increase of the relaxation time, there is indeed a strongly increasing correlation length scale, namely that of the elastic-like quadrupolar response of the material. A thorough study of this length and its implication for spontaneous strain correlations upon approaching the glass transition remains an interesting challenge for future work.


\end{document}